\documentclass[a4paper,12pt]{article}

\pdfoutput=1
\usepackage{jheppub}
\usepackage[T1]{fontenc}

\usepackage{graphicx}
\usepackage{cancel}
\usepackage{amssymb}
\usepackage{textcomp}
\usepackage{amsmath}
\usepackage{bm}
\usepackage{times}
\usepackage{epsfig}
\usepackage{epstopdf}
\usepackage{color}
\usepackage{multirow}

\def\lsim{\mathrel{\raise.3ex\hbox{$<$\kern-.75em\lower1ex\hbox{$\sim$}}}}
\def\gsim{\mathrel{\raise.3ex\hbox{$>$\kern-.75em\lower1ex\hbox{$\sim$}}}}

\def\beq{\begin{equation}}
\def\eeq{\end{equation}}
\def\be{\begin{equation}}
\def\ee{\end{equation}}
\def\bea{\begin{eqnarray}}
\def\eea{\end{eqnarray}}

\title{AMS-02 fits Dark Matter}

\author{Csaba Bal\'azs$^{\bf a}$}
\author{Tong Li$^{\bf a}$}

\emailAdd{csaba.balazs@monash.edu, tong.li@monash.edu}
\affiliation{
$^{\bf a}$  ARC Centre of Excellence for Particle Physics at the Tera-scale, School of Physics and Astronomy, Monash University, Melbourne, Victoria 3800 Australia}

\abstract{
In this work we perform a comprehensive statistical analysis of the AMS-02 electron, positron fluxes and the antiproton-to-proton ratio in the context of a simplified dark matter model.  We include known, standard astrophysical sources and a dark matter component in the cosmic ray injection spectra.  To predict the AMS-02 observables we use propagation parameters extracted from observed fluxes of heavier nuclei and the low energy part of the AMS-02 data.  We assume that the dark matter particle is a Majorana fermion coupling to third generation fermions via a spin-0 mediator, and annihilating to multiple channels at once.
The simultaneous presence of various annihilation channels provides the dark matter model with additional flexibility, and this enables us to simultaneously fit all cosmic ray spectra using a simple particle physics model and coherent astrophysical assumptions.  Our results indicate that AMS-02 observations are not only consistent with the dark matter hypothesis within the uncertainties, but adding a dark matter contribution improves the fit to the data.  Assuming, however, that dark matter is solely responsible for this improvement of the fit, it is difficult to evade the latest CMB limits in this model.
}

\begin{document}

\maketitle
\flushbottom
\newpage

\section{Introduction}

Charged cosmic rays carry a wealth of information about galactic astrophysics and possibly about new fundamental particle physics.  Deciphering this information is, however, challenging because it requires the detailed understanding the injection and propagation of cosmic rays within the Galaxy.  Fortunately, the last decade witnessed an increasing precision both in the experimental determination and the theoretical prediction of cosmic ray fluxes.  As observations became more and more precise a deviation between them and prediction became apparent in the electron and positron fluxes~\cite{Golden:1992zm, Alcaraz:2000bf, Boezio:2001ac, Grimani:2002yz, Barwick:1997ig, Beatty:2004cy, Adriani:2008zr, Delahaye:2008ua, Delahaye:2010ji, Mertsch:2010qf, Timur:2011vv, Aguilar:2002ad, Torii:2008xu, Aharonian:2008aa, Aharonian:2009ah, Ackermann:2010ij}.  The latest and most precise measurements of the electron, positron flux, antiproton-to-proton ratio, and proton flux came from the AMS-02 collaboration \cite{Accardo:2014lma,Aguilar:2014mma,Aguilar:2014fea,AMSpbarp,Aguilar:2015ooa}.  The increase of the positron spectral index and the growth of the positron fraction above 100 GeV are unexpected features of these measurements  \cite{Accardo:2014lma,Aguilar:2014mma}.

The difference between these measurements and various predictions is the subject of debate.  It may originate from unsatisfactory understanding of cosmic ray propagation, through unaccounted standard astrophysical sources (such as pulsars and/or supernova remnants), to more exotic new physics (such as dark matter annihilation)~\cite{Serpico:2011wg,Belotsky:2014nba,Mambrini:2015sia}.  Motivated by the exciting possibility that the apparent excess of cosmic electrons and positrons is due to dark matter annihilation, in this work we examine whether the AMS-02 data are consistent with a typical particle dark matter model.  First, we make a prediction for the expected background based on the propagation parameters of heavier cosmic isotopes and commonly used injection spectra.  Then we calculate the contribution of dark matter annihilation to the electron, positron and anti-proton fluxes.  Adding this to the background flux allows us to constrain the parameter space of the dark matter model.

To determine the cosmic ray background due to standard astrophysical sources we adopt the following strategy.  We assume that the relevant cosmic ray propagation parameters and injection spectra can be determined by fitting the observed fluxes and the secondary-to-primary ratios of heavier nuclei (e.g. $\rm B/C, ^{10}Be/^{9}Be$) and the low energy regions of the $e^\pm$ and $\bar{p}/p$ spectra.  Based on these fits we derive the background for the $e^\pm$ and $\bar{p}/p$ fluxes.  Then we calculate the injection spectra of $e^\pm$ and $\bar{p}$ due to dark matter annihilation.  Using the earlier determined diffusion parameters we propagate the dark matter annihilation products through the Galaxy.  This procedure ensures a consistent astrophysical treatment of cosmic rays originating from standard astrophysical sources and from dark matter. 

As particle physics description of dark matter we use the simplified model framework.  This ansatz uses minimal and general theoretical assumptions.  We consider a single dark matter particle, a Majorana fermion, that couples to standard fermions via a spin-0 mediator. 
We do not assume a specific, single annihilation final state for the dark matter particle.  Rather, more realistically and in line with minimal flavor violation~\cite{D'Ambrosio:2002ex}, we allow the dark matter particle to annihilate into the third generation quarks and the tau lepton.  The simultaneous presence of various annihilation channels provides the dark matter model with considerable flexibility, which enables us to simultaneously fit all cosmic ray spectra using a single particle physics model and coherent astrophysical assumptions.  This is one of the most important results of our work.  Beyond this outcome we also delineate the AMS-02 preferred region in the parameter space of the dark matter model.


This paper is organized as follows. In Sec.~\ref{sec:Propagation} we describe the propagation equation and injection spectra for cosmic ray in galaxy.  The values of corresponding parameters are also given.
In Sec.~\ref{sec:Models}, we briefly describe the simplified dark matter model we use.
Our numerical results are given in Sec.~\ref{sec:Results}.
Finally, in Sec.~\ref{sec:Concl} we summarize our main results.

\section{Injection and Propagation of Cosmic Rays}
\label{sec:Propagation}

Cosmic rays are energetic particles propagating within the Galaxy, and are divided into primary and secondary types \cite{1964ocr..book.....G, Blandford:1987pw, Stawarz:2009ig, Aharonian:2011da}.  Primary cosmic rays are likely to originate from powerful astrophysical processes, such as supernova explosions and pulsars.  By interacting with intergalactic matter they create secondary cosmic rays \cite{Delahaye:2008ua, Blandford:1987pw, Adriani:2008zr, Delahaye:2009gd, Nakamura:2010zzi, Aharonian:2011da}.
Propagation of charged cosmic rays within the Galaxy can be quantified by the diffusion model \cite{Ginzburg:1990sk, Schlickeiser:2002pg, Ptuskin:2005ax, Strong:2007nh}.  This model provides a mechanism to explain the retention and isotropic distribution of high energy charged particles within the Galaxy, by describing particle scattering on Galactic media, such as magnetic fields \cite{1964ocr..book.....G, Strong:2007nh, Cotta:2010ej, Aharonian:2011da}.  The spectrum of cosmic rays is modified by various energy loss mechanisms (due to interaction with the interstellar medium) and re-acceleration (due to interstellar shocks) \cite{Strong:1998pw, Strong:2007nh, Fan:2010yq}.

Cosmic ray propagation within the galactic halo is described by the transport equation \citep{Strong:2007nh}
\begin{eqnarray}
{\partial \psi\over \partial t}&=&Q(\vec{r},p)+\vec{\nabla}\cdot \left(D_{xx}\vec{\nabla}\psi-\vec{V}\psi\right)+{\partial\over \partial p}p^2D_{pp}{\partial\over \partial p}{1\over p^2}\psi \nonumber\\
&&-{\partial\over \partial p}\left[\dot{p}\psi-{p\over 3}\left(\vec{\nabla}\cdot \vec{V}\right)\psi\right]-{\psi\over \tau_f}-{\psi\over \tau_r} .
\label{propagation}
\end{eqnarray}
Here $\psi(\vec{r},t,p)$ is the density of cosmic rays per unit of total particle momentum $p$, $\vec{V}$ is the convection velocity, and $\tau_f (\tau_r)$ is the time scale for fragmentation (radioactive decay). The spatial diffusion coefficient is written in the form
\begin{eqnarray}
D_{xx}=\beta D_0 (R/R_0)^\delta ,
\end{eqnarray}
with $R$ and $\beta$ being the rigidity and particle velocity divided by light speed respectively.  The diffusion coefficient in momentum space, i.e. $D_{pp}$, is proportional to the square of the Alfven velocity $v_A$.  The height of the cylindrical diffusion halo is  $z_0$.  The above key propagation parameters can be constrained by fitting the secondary-to-primary ratios of nuclei, that is the Boron-to-Carbon ratio ($\rm B/C$) and the Beryllium ratio ($\rm ^{10}Be/^{9}Be$).  We adopt the diffusion re-acceleration model and the values of propagation parameters shown in Table \ref{tab:parameter}, determined by the $\rm B/C$ and $\rm ^{10}Be/^{9}Be$ data~\cite{Lin:2014vja}.

Each cosmic ray species is described by an equation as Eq.~(\ref{propagation}), with species specific parameters.  The source term of cosmic ray species $i$ can be generally described by the product of spatial distribution and injection spectrum functions
\begin{eqnarray}
Q_i(\vec{r},p)=f(r,z)q_i(p) .
\end{eqnarray}
For the spatial distribution of the injected primary cosmic rays we use the following supernova remnants distribution
\begin{eqnarray}
f(r,z)=f_0\left({r\over r_\odot}\right)^a{\rm exp}\left(-b \ {r-r_\odot\over r_\odot}\right){\rm exp}\left(-{|z|\over z_s}\right),
\label{snr}
\end{eqnarray}
where the distance between the Sun and the Galactic center is $r_\odot=8.5 \ {\rm kpc}$, the height of the Galactic disk is $z_s=0.2 \ {\rm kpc}$, and the two parameters $a$ and $b$ are taken to be 1.25 and 3.56, respectively. The normalization parameter $f_0$ is determined by the EGRET gamma ray data~\cite{Jin:2014ica}. We assume the following power law with one break for the injection spectra of various nuclei
\begin{eqnarray}
q_i&\propto& \left\{
                \begin{array}{ll}
                  \left(R/R_{\rm br}^p\right)^{-\nu_1}, & R\leq R_{\rm br}^p \\
                  \left(R/R_{\rm br}^p\right)^{-\nu_2}, & R> R_{\rm br}^p
                \end{array}
              \right. \ \ \ {\rm nuclei},
\end{eqnarray}
and two breaks for primary electrons, i.e. $R_{\rm br1}^e,R_{\rm br2}^e$ with $\gamma_1,\gamma_2,\gamma_3$ being the power law indexs.

Following the approach in Ref.~\cite{Lin:2014vja} we adopt a scale factor $c_{e^+}=3.1$ to take into account the uncertainty in the calculation of the secondary fluxes from proton-proton collision cross section and enhancement factor from heavier nuclei. It is introduced to rescale the calculated secondary flux to fit the data. The corresponding injection parameters can be determined by fitting the AMS-02 proton, electron, and positron data.  We adopt injection parameters obtained by such a fit in Ref.~\cite{Lin:2014vja}.  The values of these injection parameters are shown in Table \ref{tab:parameter}.

We use the Fisk potential $\phi_i \ (i = e^-, e^+, p, \bar{p})$, relating the local interstellar fluxes to the one measured at the top of the atmosphere, to account for the solar modulation effect.  We treat $\phi_i$ as species specific nuisance parameters.  Their best fit values are shown in Table \ref{tab:parameter}.  Since solar modulation affects the observed fluxes only below 10 GeV, the values of these parameters have no effect on our conclusions drawn about the dark matter contribution.

\begin{table}[h]
\begin{center}
\resizebox{16cm}{!} {
\begin{tabular}{|c|c|c|c|c|c|c|c|c|c|c|}
        \hline
        propagation & value && nucleon injection & value && electron injection & value && solar modulation & value\\
        \hline
        $D_0 \ (10^{28} \ {\rm cm}^2 \ {\rm s}^{-1})$ & 6.58 && $\nu_1$ & 1.811 && $\gamma_1$ & 1.463 && $\phi_{e^-} \ ({\rm MV})$ & 1550\\
        \hline
        $\delta$ & 0.33 && $\nu_2$ & 2.402 && $\gamma_2$ & 2.977 && $\phi_{e^+} \ ({\rm MV})$ & 1800\\
        \hline
        $R_0 \ ({\rm GV})$ & 4 && $R_{\rm br}^p$ \ ({\rm GV}) & 12.88 && $\gamma_3$ & 2.604 && $\phi_{p} \ ({\rm MV})$ & 518\\
        \hline
        $v_A \ ({\rm km} \ {\rm s}^{-1})$ & 37.8 && $A_p$ (see caption) & 4.613 && $R^e_{\rm br1} \ ({\rm GV})$ & 2.858 && $\phi_{\bar{p}} \ ({\rm MV})$ & 0\\
        \hline
        $z_0$ \ ({\rm kpc}) & 4.7 &&    $-$    & $-$ && $R^e_{\rm br2} \ ({\rm GV})$ & 68.865 && $-$ & $-$ \\
        \hline
        $-$ & $-$ &&    $-$    & $-$ && $A_e$ (see caption) & 1.585 && $-$ & $-$ \\
        \hline
\end{tabular}}
\end{center}
\caption{Parameters of propagation, nucleon/electron injection and solar modulation and their values adopted in our numerical analysis.  The proton (electron) flux is normalized to  $A_p$ ($A_e$) at 100 (25) GeV in the units of $10^{-9} \ {\rm cm}^{-2} \ {\rm s}^{-1} \ {\rm sr}^{-1} \ {\rm MeV}^{-1}$.}
\label{tab:parameter}
\end{table}

\section{The Dark Matter Model}
\label{sec:Models}

In this section, we describe the particle physics model we use to demonstrate that the AMS-02 data can be explained by dark matter annihilation.
In the recent literature it was shown that Majorana fermions are one of the most plausible dark matter candidates \cite{Balazs:2014jla, arXiv:1407.1859, arXiv:1408.2223, arXiv:1409.5776, arXiv:1501.03164, arXiv:1503.01500, Balazs:2015boa, arXiv:1507.02288}.  Inspired by this, we assume that dark matter is composed of Majorana fermion particles, which we denote by $\chi$.  Motivated by the Higgs portal mechanism, we assume that the dark matter particle couples to standard fermions via a spin-0 mediator, that we denote by $S$ \cite{Higgsportal,deS.Pires:2010fu}.  We cast the dark matter to mediator coupling in the form
\begin{eqnarray}
{\cal L}_\chi \supset \frac{i\lambda_\chi}{2}\bar{\chi}\gamma_5\chi S.
\label{eq:interaction0}
\end{eqnarray}
Coupling between the dark matter and mediator is fixed to $\lambda_\chi=1$.  (This choice effectively absorbs $\lambda_\chi$ into the mediator-standard model couplings.)  Coupling between the mediator and standard model fermions $f$ is given by
\begin{eqnarray}
{\cal L}_S \supset \lambda_f \bar{f}f S.
\label{eq:intercation1}
\end{eqnarray}
We assume that $S$ only couples to third generation fermions, consistently with minimal flavor violation, i.e. $f=b,t,\tau$ \cite{D'Ambrosio:2002ex}.
For simplicity we do not consider dark matter annihilation into a pair of $S$ particles.
With the interactions defined by Eqs.~(\ref{eq:interaction0}) and (\ref{eq:intercation1}) dark matter annihilation is not velocity suppressed~\cite{Kumar:2013iva}.  At the same time the dark matter-nucleon elastic scattering cross section is spin-independent (SI) and momentum suppressed.

Under the above assumptions the dark matter model is described by the following parameters:
\begin{eqnarray}
P = \left\{ m_\chi, m_S, \lambda_b, \lambda_t, \lambda_\tau \right\} .
\end{eqnarray}
The scan ranges for these parameters are
\begin{eqnarray}
1 \ {\rm TeV} < m_\chi < 10 \ {\rm TeV}, ~~~~~~
1 \ {\rm GeV} < m_S < 1 \ {\rm TeV}, ~~~~~~
10^{-4} < \lambda_b,\lambda_\tau,\lambda_t < 10^5.
\end{eqnarray}
The potentially large values of the above effective couplings can only be understood in an underlying theory.  They may include the effect of large but renormalizable perturbative couplings, large loop contributions from vector-like matter, resonant or Sommerfeld enhancements, or the combination of more than one such a factor \cite{Sommerfeld}.


The source term arising from dark matter annihilation contributing to the cosmic ray species $i$ is given by
\begin{eqnarray}
Q_i^\chi(r,p)=\frac{\rho_\chi^2(r)\langle \sigma v\rangle}{2 m_\chi^2}\left(\sum_f B_f \frac{dN_i^f}{dE}\right),
\label{dmsource}
\end{eqnarray}
where $\langle \sigma v\rangle$ is the velocity averaged dark matter annihilation cross section, $B_f=\langle \sigma v\rangle_f/ \langle \sigma v\rangle$ is the annihilation fraction into the $f{\bar f}$ final state, and $dN_i^f/ dE$ is the energy spectrum of cosmic ray particle $i$ produced in the annihilation channel into $f\bar{f}$.  In the parenthesis on the right hand side the total differential yield is the $B_f$ weighted sum of the partial differential yields into specific final states.  The sum includes contributions from all the third generation charged fermions ($b,t,\tau$).  AMS-02 plays an important role in constraining the coupling of the mediator to these fermions since $B_f$ directly depends on these couplings.

We use a generalized Navarro-Frenk-White (NFW) profile to describe dark matter spatial distribution within the Galaxy~\cite{NFW}
\begin{eqnarray}
\rho_\chi(r)=\rho_0\frac{(r/r_s)^{-\gamma}}{(1+r/r_s)^{3-\gamma}}.
\end{eqnarray}
Here
the normalization coefficient is $\rho_0=0.26 \ {\rm GeV/cm^3}$ and
the radius of the galactic diffusion disk is $r_s=20$ kpc.
We fix the inner slope of the halo profile to $\gamma=1$.


\section{Results}
\label{sec:Results}

As discussed in Sec.~\ref{sec:Propagation}, the propagation and injection parameters of cosmic rays are determined by fitting the $\rm B/C$ and $\rm ^{10}Be/^{9}Be$ data and recent charged cosmic ray data from AMS-02, respectively \cite{Lin:2014vja}.  The parameters in Table~\ref{tab:parameter} thus imply prediction for cosmic ray measurements inferred from standard astrophysical sources.  One can investigate the constraint on extra sources, such as dark matter, based on this fiducial astrophysical background.

To this end the Lagrangian of the dark matter model described in the previous section was coded in FeynRules~\cite{FR}.  Using model files generated by FeynRules, the annihilation fraction $B_f$ and differential yields $dN_i^f/dE$ in Eq.~(\ref{dmsource}) were calculated by a modified version of micrOmegas 3.6.9~\cite{MO}.  These dark matter model dependent variables were then input into the public code Galprop v54~\cite{Strong:1998pw, Moskalenko:2001ya, Strong:2001fu, Moskalenko:2002yx, Ptuskin:2005ax} to ensure that near Earth cosmic ray fluxes from dark matter annihilation and background spectra obtained in a consistent way.

The calculated cosmic ray fluxes, together with the measured spectral data points, were entered in a composite likelihood function, defined as
\begin{eqnarray}
- 2\ln{\cal L}= \sum_i {(f_i^{\rm th}-f_i^{\rm exp})^2\over \sigma_i^2} .
\label{sum}
\end{eqnarray}
Here $f_i^{\rm th}$ are the theoretical predictions and $f_i^{\rm exp}$ are the corresponding central value of the experimental data.   The uncertainty $\sigma_i$ combines the theoretical and experimental uncertainties in quadrature.  We stipulate a 50\% uncertainty of the theoretical prediction of electron flux, positron flux and antiproton-proton ratio according to the estimates of Refs.~\cite{Trotta:2010mx, Auchettl:2011wi, Yuan:2014pka, Giesen:2015ufa}.  This uncertainty takes into account, amongst other, the uncertainty related to the fixed propagation parameters.
The sum in Eq.~(\ref{sum}) runs over all the AMS cosmic ray spectral data points: the electron flux (73 points), positron flux (72 points) and antiproton-proton ratio (30 points).  We do not include the AMS-02 positron fraction data in the likelihood function; consequently the theoretical positron fraction flux is a prediction in our framework.

Including observables from dark matter abundance, direct detection, or collider production in the likelihood function would not change its value significantly.  We found that in the parameter region that dark matter annihilation can appreciably contribute to the charged cosmic ray fluxes the self-annihilation rate is high enough to decrease dark matter abundance below the observed level.  In this case, assuming that $\chi$ is just a component of dark matter, the likelihood is not affected by abundance.  Dark matter direct detection is impaired by momentum suppressed $\chi$-nucleon elastic scattering cross section and the very high mass of $\chi$.  As for the Large Hadron Collider (LHC), in the relevant parameter region $\chi$ particles are too heavy to produce in significant numbers via 14 TeV proton-proton collisions.

\begin{figure}[t]
\begin{center}
\includegraphics[scale=1,width=7cm]{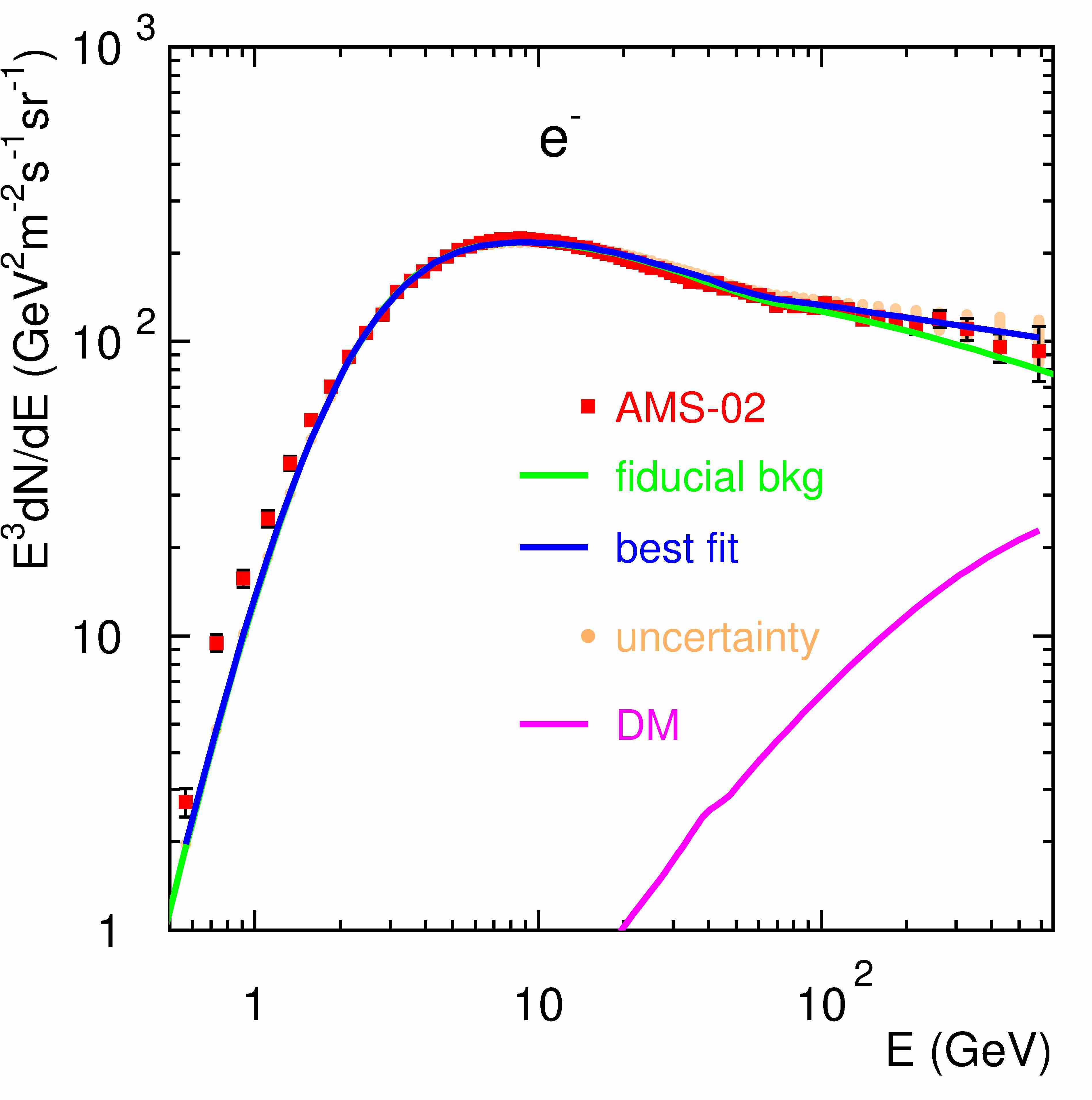}
\includegraphics[scale=1,width=7cm]{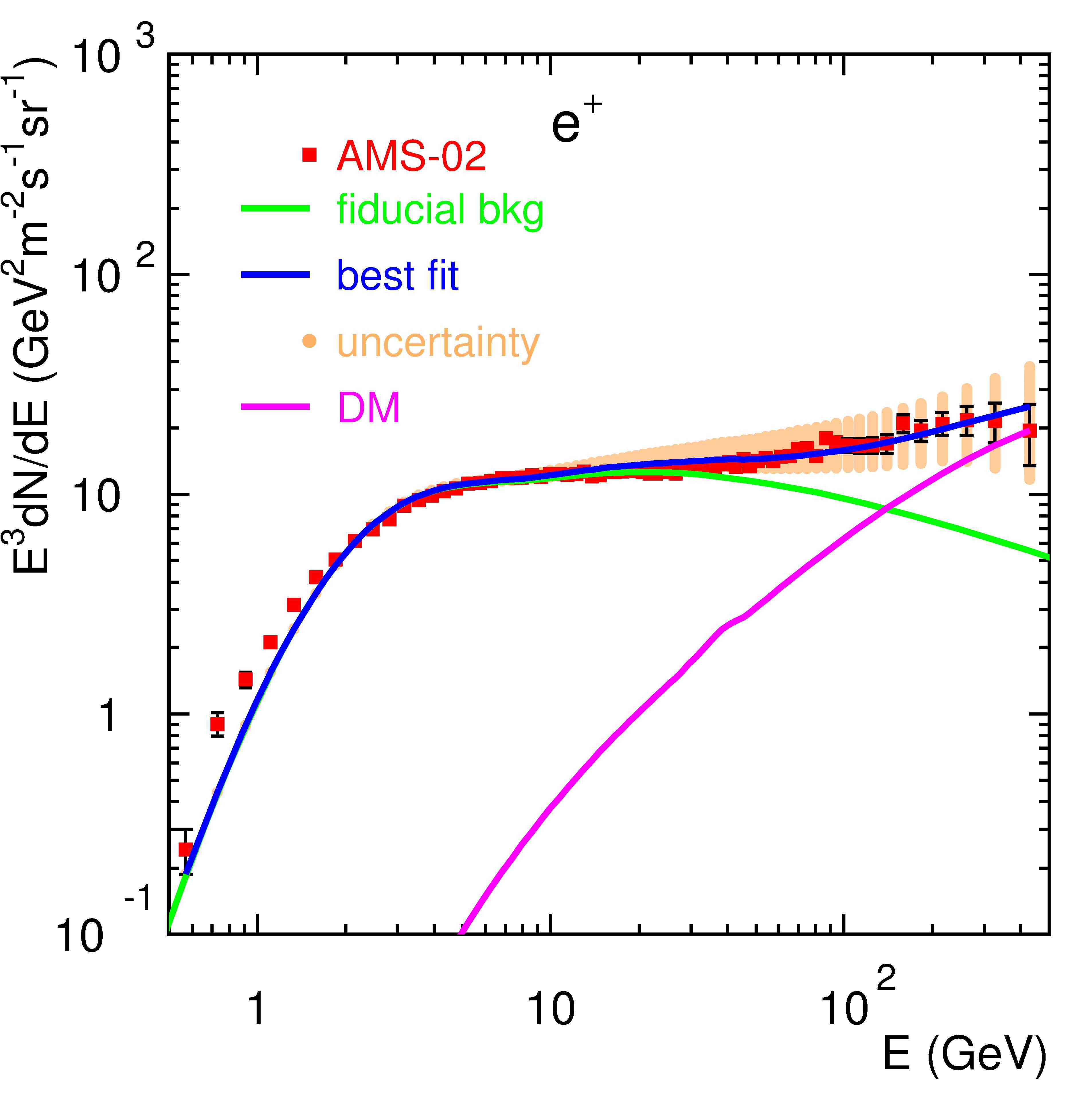}\\
\includegraphics[scale=1,width=7cm]{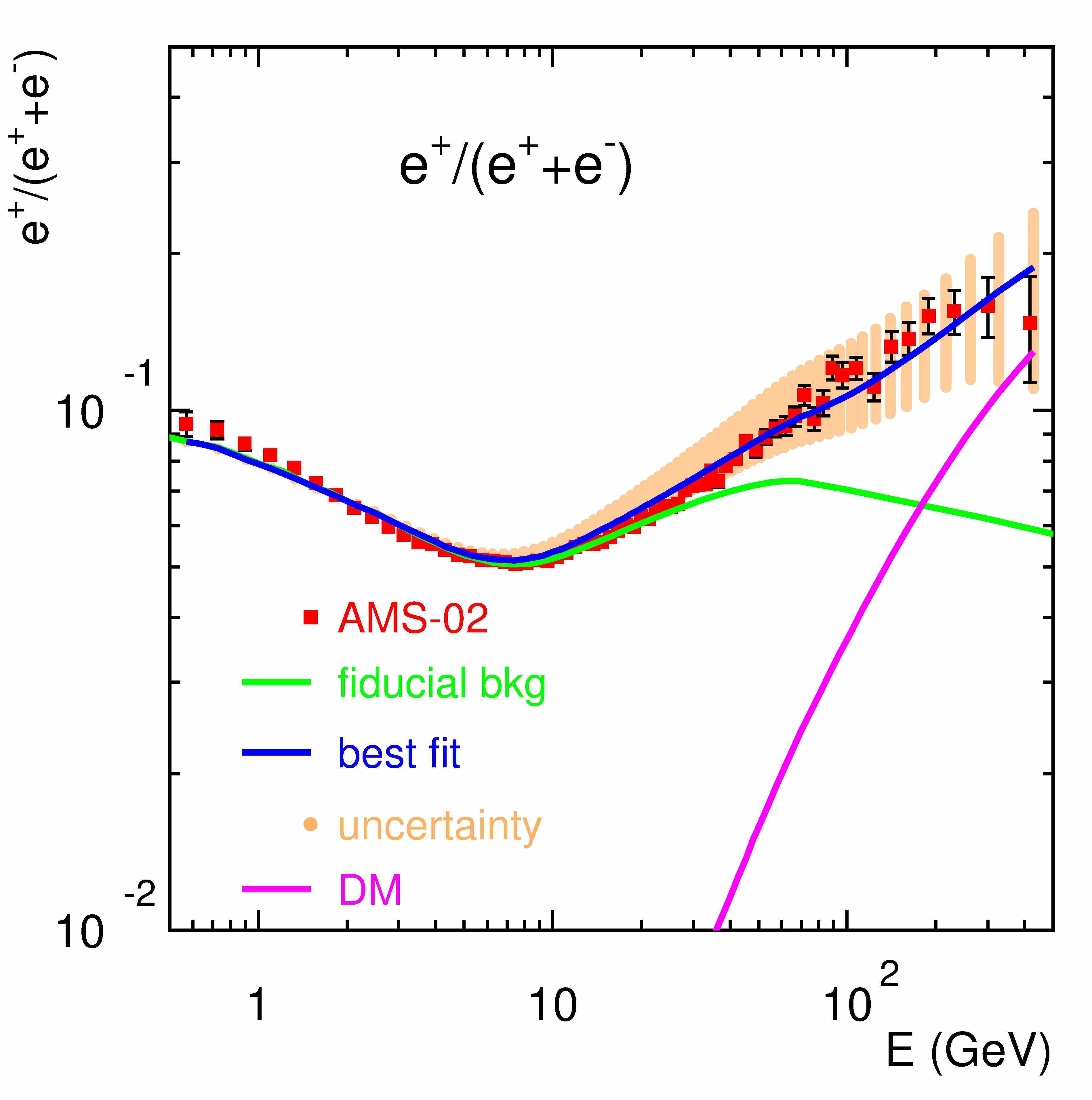}
\includegraphics[scale=1,width=7cm]{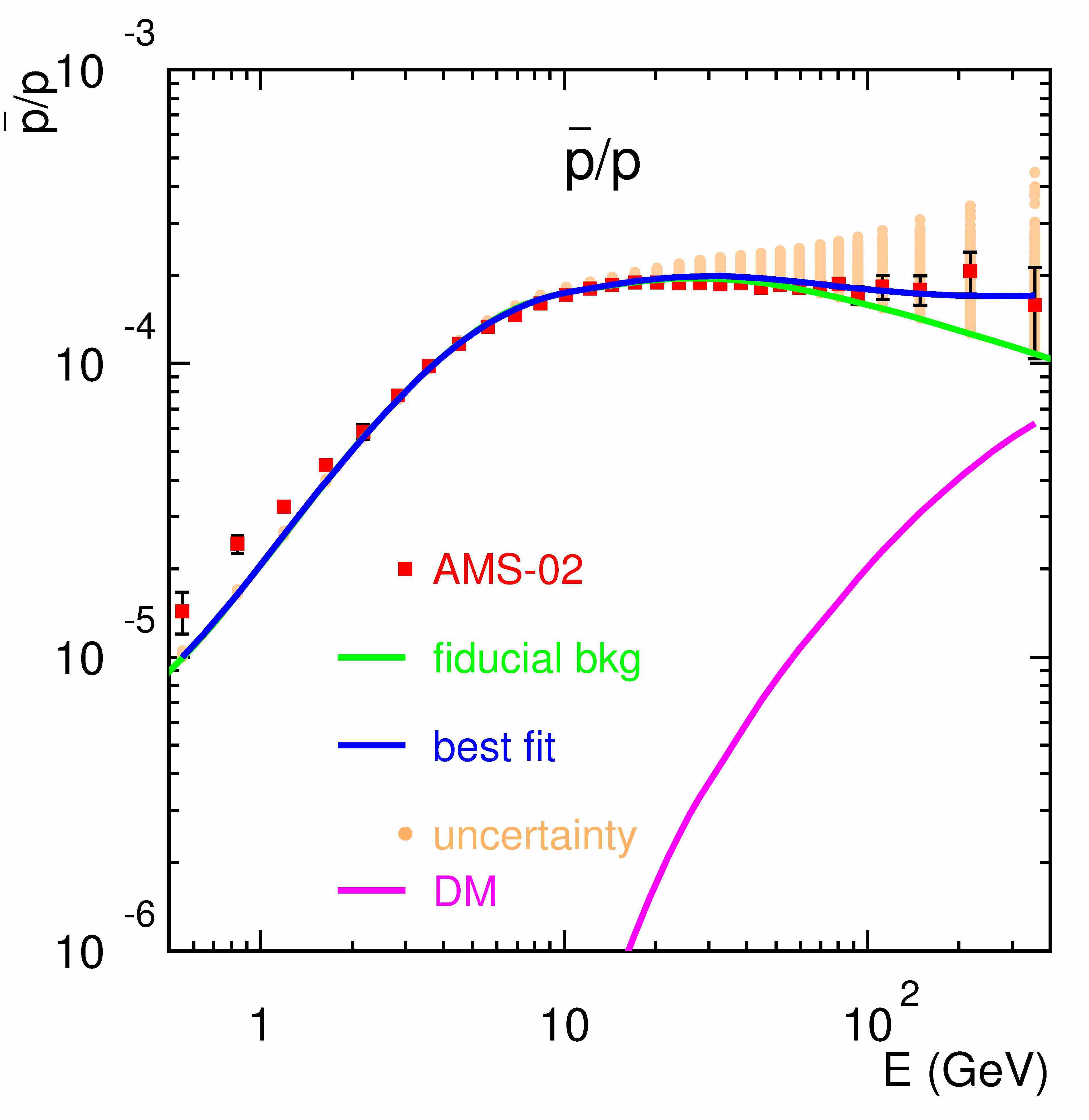}
\end{center}
\caption{Electron flux, positron flux, positron fraction, and antiproton-to-proton ratio observed by AMS-02 (red dots and dark error bars).  The blue solid line shows the prediction of the total cosmic ray flux with dark matter parameter values that best fit the AMS-02 data. The total predicted flux is the sum of the background flux (green solid line) and the dark matter contribution.  Orange dots indicate the 95\% confidence region of the prediction. The magenta line is the flux from dark matter at the best fit point.
}
\label{fig:fit}
\end{figure}

FIG. \ref{fig:fit} shows our main results: AMS-02 cosmic ray flux observations are consistent with the dark matter hypothesis within the uncertainties.  The four frames display the various cosmic ray fluxes AMS-02 observed: electron flux, positron flux, positron fraction, and antiproton-to-proton ratio.  AMS-02 central value measurements are shown by red dots and dark error bars indicate their uncertainty.  The green solid line, on each frame, is obtained using the parameters shown in Table~\ref{tab:parameter} and displays the predicted background flux originating from standard astrophysical sources.  The blue solid line shows the prediction of the total cosmic ray flux with dark matter parameter values that best fit the AMS-02 data.  The blue curve is the sum of the background flux (green curve) and the dark matter contribution at the best fit point (magenta curve).  A series of orange colored dots (forming vertical bars) indicate the theoretical uncertainty of the dark matter prediction given by the 95\% confidence region of dark matter model parameters.

As the plots show adding a dark matter contribution to the background flux yields a better fit to the AMS-02 data.  As expected, the electron flux is hardly changed by the dark matter contribution, while the latter somewhat improves the agreement between the theoretical prediction and the antiproton-to-proton ratio data.  This indicates that the dark matter model is consistent with these data.
The fit to the positron data is noticeably improved that justifies the addition of the dark matter component.
%
Our likelihood function used to extract the best-fit dark matter parameters does not include the positron fraction data, that is the dark matter model parameters are not fit to the $e^+/(e^+ +e^-)$ fraction.  Rather, after we extract the best fit dark matter model parameters, we calculate the positron fraction using the best fit parameters.  As shown by the blue curve the $e^+/(e^+ +e^-)$ fraction data and the best fit (obtained without this data) agree very well.  This is an important cross check of the internal consistency of the dark matter model and our parameter extraction procedure.

\begin{figure}[t]
\begin{center}
\includegraphics[scale=1,width=7.5cm]{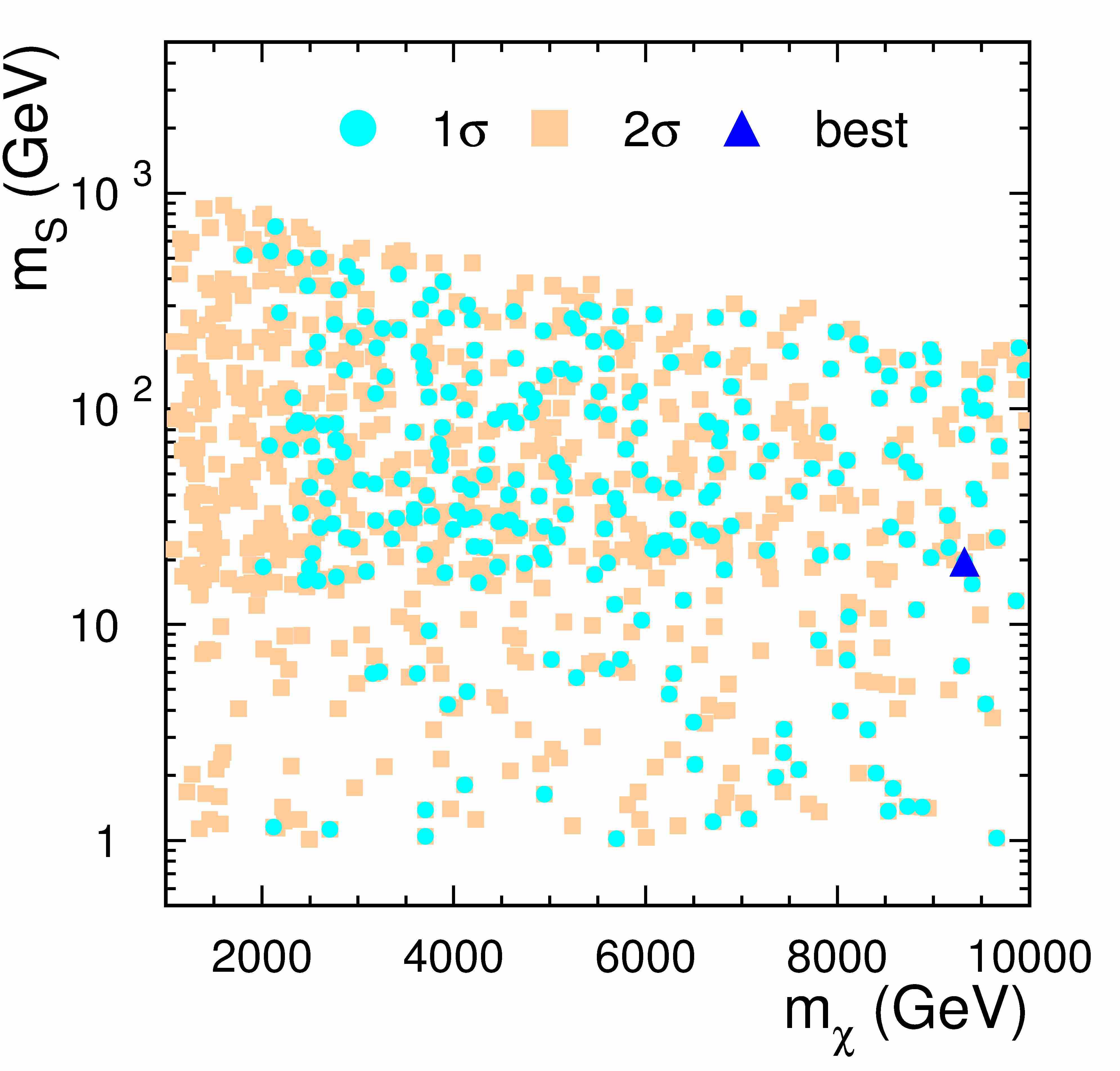}
\includegraphics[scale=1,width=7.5cm]{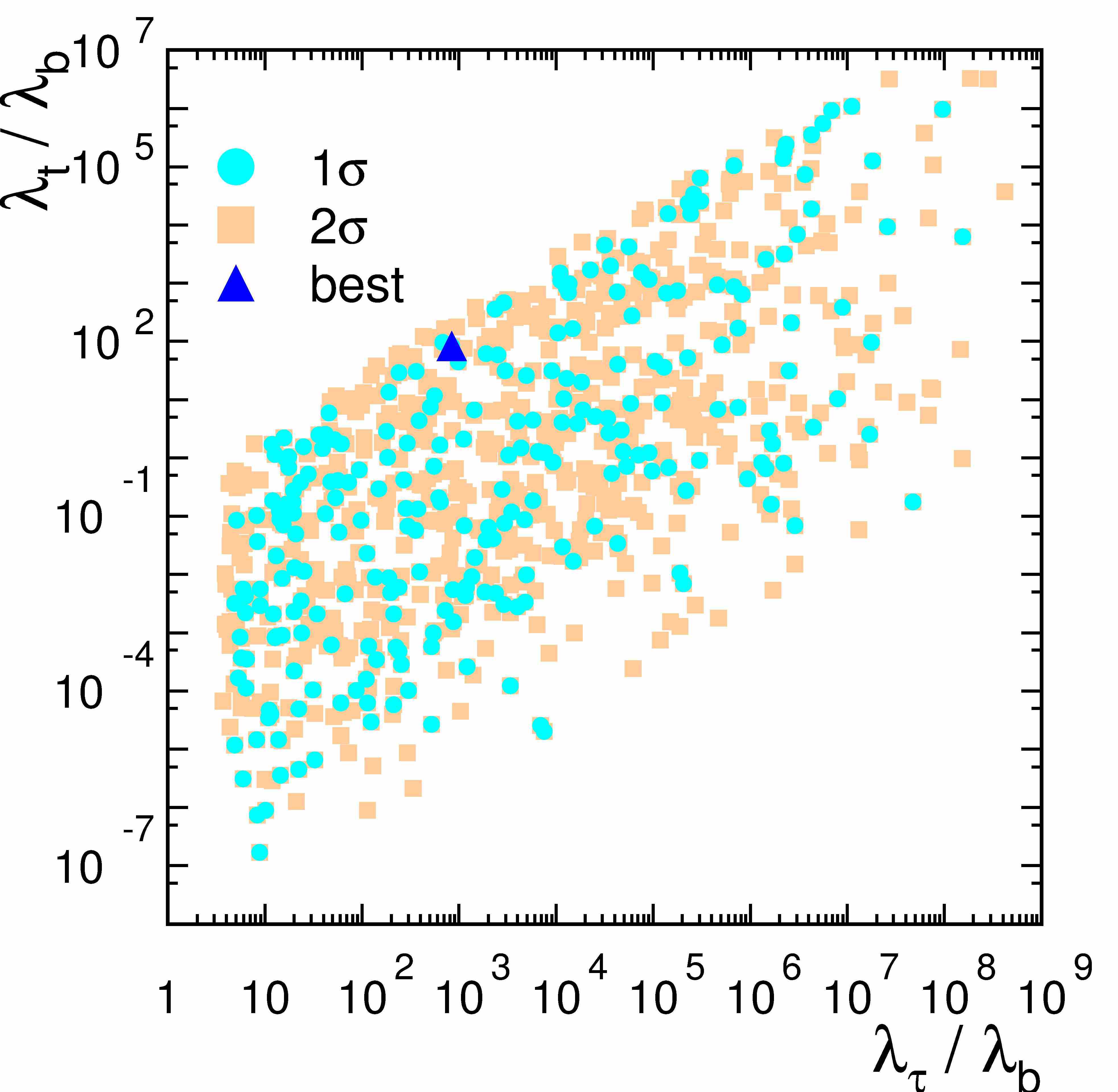}
\includegraphics[scale=1,width=7.5cm]{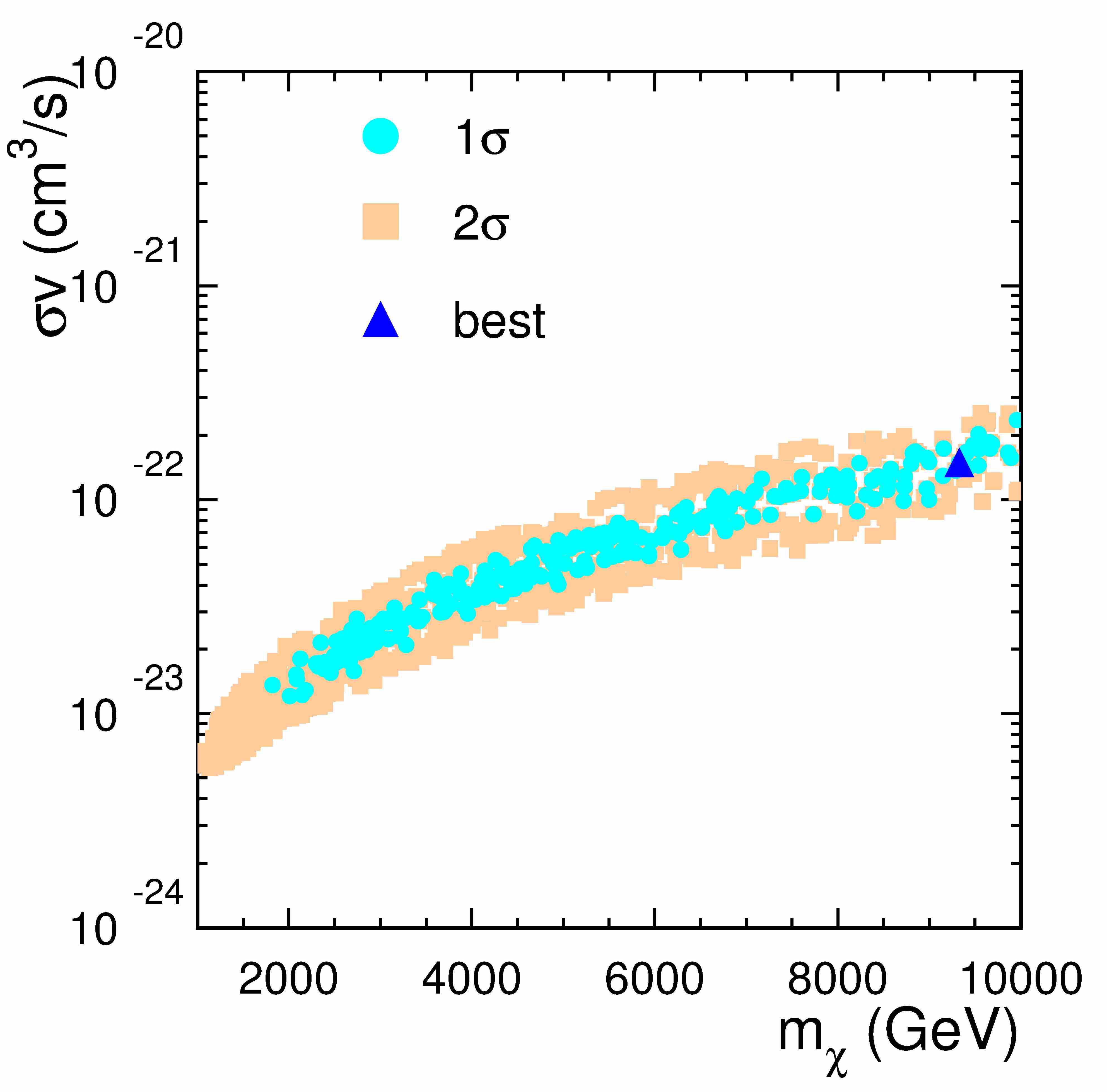}
\end{center}
\caption{The AMS-02 favored region of masses (top left, $m_S$ vs. $m_\chi$), couplings (top right, $\lambda_t/\lambda_b$ vs. $\lambda_\tau/\lambda_b$), and cross sections (bottom, $\sigma v$ vs. $m_\chi$) in the simplified dark matter model we consider.  The solid circles and squares estimate 68\% and 95\% confidence regions, respectively.  The best fit point is indicated by a triangle.}
\label{fig:region}
\end{figure}

The top frames of FIG.~\ref{fig:region} show the regions of the dark matter parameter space preferred by the AMS-02 data.  Solid circles and squares denote the estimated 68\% and 95\% confidence regions, respectively.  The favored mass of the dark matter particle is heavier than 2 TeV (at about 68\% C.L.) with best fit point indicating an 9.3 TeV dark matter mass.  The AMS-02 data favor a spin-0 mediator mass in the region of 1--700 GeV (at about 68\% C.L.).

For the mediator-SM fermion couplings the favored region indicates that the tau lepton coupling $\lambda_\tau$ is generally larger than quark couplings $\lambda_b, \lambda_t$, being 1000 (10) times larger than $\lambda_b$ ($\lambda_t$) at the best fit point.  This trend is governed by the electron and positron data fit: dark matter annihilations should produce mostly leptons to explain the difference between the astrophysical background and the AMS-02 data at high energies.
The antiproton-to-proton ratio data, on the other hand, require the moderate presence of either bottom or top quarks in the final state.  Hence the diagonal shape of the estimated 68\% and 95\% C.L. regions on the right hand frame of FIG.~\ref{fig:region}.  The best fit point favors coupling values for which $\lambda_\tau \sim 10 \lambda_t \sim 1000 \lambda_b$.

The bottom frame of FIG.~\ref{fig:region} shows that the AMS-02 data require an effective dark matter annihilation cross section in the region of $1 \times 10^{-23}$ -- $2 \times 10^{-22}$ ($5 \times 10^{-24}$ -- $3 \times 10^{-22}$) ${\rm cm}^3/{\rm s}$ at about 68 (95) \% C.L.
%
An effective cross section so much higher than the standard thermal rate could indicate the non-thermal origin of self-annihilating dark matter particles responsible for AMS-02 \cite{arXiv:1204.2795, arXiv:1307.2453, arXiv:1502.05406}.
Alternatively, the positron ray flux might receive a boost from dark matter substructure, such as over dense clumps, clouds, or disks which would allow for a reduced annihilation rate \cite{arXiv:1310.1915, arXiv:0905.2736, arXiv:0812.3202, arXiv:0809.1523, arXiv:0805.1244, arXiv:0803.2714, arXiv:0902.0009, arXiv:0902.4001, arXiv:0906.5348, McCullough:2013jma}.

According to Ref. \cite{Elor:2015bho} a 1-10 TeV dark matter particle with an annihilation cross section of $\sigma v \sim 10^{-23}-10^{-22} ~{\rm cm^3/s}$, and dominant final state of $\tau^+ \tau^-$ or $b {\bar b}$, is excluded by Planck and by Fermi-LAT gamma ray bounds from dwarf satellite galaxies.  Since the annihilation rate at the recombination time places a (particle physics) model independent limit on the present day annihilation rate, either of these limits are hard to evade.  Sommerfeld enhancement does not alleviate the problem, since the average relative velocity of scattering dark matter particles at the time of CMB is lower than the present day one.  Uncertainties in the relevant astrophysical measurements, such as in the power injected into the CMB or the Fermi-LAT statistical/systematic errors, do not seem to leave enough room for the high dark matter annihilation cross section required to account for AMS-02.
The most straightforward way to evade the Planck and Fermi-LAT limits appears to be including a standard, but presently unanticipated, astrophysical contribution to explain the AMS-02 measurements.  With such additional contribution the dark matter annihilation cross section can be lowered and the model be made consistent with all data.

%

\section{Conclusions}
\label{sec:Concl}

In this work we examined the plausibility of dark matter annihilation contributing to the recent AMS-02 data, the electron, positron fluxes and antiproton-to-proton ratio.
On the top of the standard astrophysical cosmic ray flux prediction we included a dark matter component.  Our choice of the dark matter model was a Majorana fermion coupling to third generation fermions via a spin-0 mediator.  The initial flux from standard astrophysical sources and dark matter annihilation were propagated through the Galaxy using the same set of diffusion parameters.  The latter were determined by fitting the cosmic ray fluxes of heavier elements and the low energy regions of the AMS-02 data.

We have shown that not only AMS-02 observations are consistent with the dark matter hypothesis within the uncertainties, but adding a dark matter contribution to the background flux yields a better fit to the data.
We also estimated the most plausible parameter regions of the dark matter parameter space in light of AMS-02.
The observations prefer a dark matter (mediator) mass in the 2--10 TeV (1--700 GeV) region at about 68\% confidence level.
The data also favor a dominant tau lepton--dark matter coupling $\lambda_\tau$, about ten times larger than top quark--dark matter coupling $\lambda_t$ at the best fit point.  The antiproton-to-proton ratio data require that dark matter annihilation to quarks is dominated by either the top or the bottom final state with a slight preference for the latter.

At the meantime we found it to be difficult to evade the CMB and Fermi-LAT gamma ray limits in this model due to the high annihilation cross section.  With additional contribution to the positron spectrum from standard, but presently unknown, astrophysics this cross section can be lowered and the model be made consistent with all data.

\acknowledgments
We thank Xiao-Jun Bi and Qiang Yuan for helping with Galprop.
This work in part was supported by the ARC Centre of Excellence for Particle Physics at the Terascale.
%
%
The National Computational Infrastructure (NCI), the Southern Hemisphere's fastest supercomputer, is also gratefully acknowledged.

\appendix



\begin{thebibliography}{100}


\bibitem{Golden:1992zm}
  R.~L.~Golden {\it et al.},
  Astrophys.\ J.\  {\bf 436}, 769 (1994).


\bibitem{Alcaraz:2000bf}
  J.~Alcaraz {\it et al.} [AMS Collaboration],
  Phys.\ Lett.\ B {\bf 484}, 10 (2000)
  [Phys.\ Lett.\ B {\bf 495}, 440 (2000)].


\bibitem{Boezio:2001ac}
  M.~Boezio {\it et al.} [WiZard/CAPRICE Collaboration],
  Astrophys.\ J.\  {\bf 561}, 787 (2001)
  [astro-ph/0103513].


\bibitem{Grimani:2002yz}
  C.~Grimani {\it et al.},
  Astron.\ Astrophys.\  {\bf 392}, 287 (2002).


\bibitem{Barwick:1997ig}
  S.~W.~Barwick {\it et al.} [HEAT Collaboration],
  Astrophys.\ J.\  {\bf 482}, L191 (1997)
  [astro-ph/9703192].


\bibitem{Beatty:2004cy}
  J.~J.~Beatty {\it et al.},
  Phys.\ Rev.\ Lett.\  {\bf 93}, 241102 (2004)
  [astro-ph/0412230].


\bibitem{Adriani:2008zr}
  O.~Adriani {\it et al.} [PAMELA Collaboration],
  Nature {\bf 458}, 607 (2009)
  [arXiv:0810.4995 [astro-ph]].


\bibitem{Delahaye:2008ua}
  T.~Delahaye, F.~Donato, N.~Fornengo, J.~Lavalle, R.~Lineros, P.~Salati and R.~Taillet,
  Astron.\ Astrophys.\  {\bf 501}, 821 (2009)
  [arXiv:0809.5268 [astro-ph]].


\bibitem{Delahaye:2010ji}
  T.~Delahaye, J.~Lavalle, R.~Lineros, F.~Donato and N.~Fornengo,
  Astron.\ Astrophys.\  {\bf 524}, A51 (2010)
  [arXiv:1002.1910 [astro-ph.HE]].


\bibitem{Mertsch:2010qf}
  P.~Mertsch,
  arXiv:1012.4239 [astro-ph.HE].


\bibitem{Timur:2011vv}
  T.~Delahaye, A.~Fiasson, M.~Pohl and P.~Salati,
  Astron.\ Astrophys.\  {\bf 531}, A37 (2011)
  [arXiv:1102.0744 [astro-ph.HE]].


\bibitem{Aguilar:2002ad}
  M.~Aguilar {\it et al.} [AMS Collaboration],
  Phys.\ Rept.\  {\bf 366}, 331 (2002)
  [Phys.\ Rept.\  {\bf 380}, 97 (2003)].


\bibitem{Torii:2008xu}
  S.~Torii {\it et al.} [PPB-BETS Collaboration],
  arXiv:0809.0760 [astro-ph].


\bibitem{Aharonian:2008aa}
  F.~Aharonian {\it et al.} [HESS Collaboration],
  Phys.\ Rev.\ Lett.\  {\bf 101}, 261104 (2008)
  [arXiv:0811.3894 [astro-ph]].


\bibitem{Aharonian:2009ah}
  F.~Aharonian {\it et al.} [HESS Collaboration],
  Astron.\ Astrophys.\  {\bf 508}, 561 (2009)
  [arXiv:0905.0105 [astro-ph.HE]].


\bibitem{Ackermann:2010ij}
  M.~Ackermann {\it et al.} [Fermi-LAT Collaboration],
  Phys.\ Rev.\ D {\bf 82}, 092004 (2010)
  [arXiv:1008.3999 [astro-ph.HE]].


\bibitem{Accardo:2014lma}
L.~Accardo {\it et al.} [AMS Collaboration],
Phys.\ Rev.\ Lett.\  {\bf 113}, 121101  (2014).

\bibitem{Aguilar:2014mma}
  M.~Aguilar {\it et al.} [AMS Collaboration],
  Phys.\ Rev.\ Lett.\  {\bf 113}, 121102 (2014).

\bibitem{Aguilar:2014fea}
M.~Aguilar {\it et al.} [AMS Collaboration],
Phys.\ Rev.\ Lett.\  {\bf 113}, 221102 (2014).
doi:10.1103/PhysRevLett.113.221102

\bibitem{AMSpbarp}
AMS-02 Collaboration, talks at the ``AMS Days at CERN'', April 15-17, 2015.

\bibitem{Aguilar:2015ooa}
M.~Aguilar {\it et al.} [AMS Collaboration],
Phys.\ Rev.\ Lett.\  {\bf 114}, 171103 (2015).
doi:10.1103/PhysRevLett.114.171103




\bibitem{Serpico:2011wg}
  P.~D.~Serpico,
  Astropart.\ Phys.\  {\bf 39-40}, 2 (2012)
  [arXiv:1108.4827 [astro-ph.HE]].

\bibitem{Belotsky:2014nba}
  K.~Belotsky, M.~Khlopov and M.~Laletin,
  arXiv:1411.3657 [hep-ph].

\bibitem{Mambrini:2015sia}
  Y.~Mambrini, S.~Profumo and F.~S.~Queiroz,
  arXiv:1508.06635 [hep-ph].

\bibitem{D'Ambrosio:2002ex}
  G.~D'Ambrosio, G.~F.~Giudice, G.~Isidori and A.~Strumia,
  Nucl.\ Phys.\ B {\bf 645}, 155 (2002)
  [hep-ph/0207036].

\bibitem{1964ocr..book.....G}
Ginzburg, V. L., Syrovatskii, S. I., The Origin of Cosmic Rays (Macmillan, New York, 1964)

\bibitem{Blandford:1987pw}
R.~Blandford and D.~Eichler,
Phys.\ Rept.\  {\bf 154}, 1 (1987).


\bibitem{Stawarz:2009ig}
  L.~Stawarz, V.~Petrosian and R.~D.~Blandford,
  Astrophys.\ J.\  {\bf 710}, 236 (2010)
  [arXiv:0908.1094 [astro-ph.GA]].


\bibitem{Aharonian:2011da}
  F.~Aharonian, A.~Bykov, E.~Parizot, V.~Ptuskin and A.~Watson,
  Space Sci.\ Rev.\  {\bf 166}, 97 (2012)
  [arXiv:1105.0131 [astro-ph.HE]].


\bibitem{Nakamura:2010zzi}
  K.~Nakamura {\it et al.} [Particle Data Group Collaboration],
  J.\ Phys.\ G {\bf 37}, 075021 (2010).


\bibitem{Delahaye:2009gd}
  T.~Delahaye, P.~Brun, F.~Donato, N.~Fornengo, J.~Lavalle, R.~Lineros, R.~Taillet and P.~Salati,
  arXiv:0905.2144 [hep-ph].


\bibitem{Ginzburg:1990sk}
  V.~L.~Ginzburg, V.~A.~Dogiel, V.~S.~Berezinsky, S.~V.~Bulanov and V.~S.~Ptuskin,
  Amsterdam, Netherlands: North-Holland (1990) 534 p


\bibitem{Schlickeiser:2002pg}
  R.~Schlickeiser,
  Berlin, Germany: Springer (2002) 519 p


\bibitem{Ptuskin:2005ax}
  V.~S.~Ptuskin, I.~V.~Moskalenko, F.~C.~Jones, A.~W.~Strong and V.~N.~Zirakashvili,
  Astrophys.\ J.\  {\bf 642}, 902 (2006)
  [astro-ph/0510335].


\bibitem{Strong:2007nh}
  A.~W.~Strong, I.~V.~Moskalenko and V.~S.~Ptuskin,
  Ann.\ Rev.\ Nucl.\ Part.\ Sci.\  {\bf 57}, 285 (2007)
  [astro-ph/0701517].


\bibitem{Cotta:2010ej}
  R.~C.~Cotta, J.~A.~Conley, J.~S.~Gainer, J.~L.~Hewett and T.~G.~Rizzo,
  JHEP {\bf 1101}, 064 (2011)
  [arXiv:1007.5520 [hep-ph]].


\bibitem{Strong:1998pw}
  A.~W.~Strong and I.~V.~Moskalenko,
  Astrophys.\ J.\  {\bf 509}, 212 (1998)
  [astro-ph/9807150].


\bibitem{Fan:2010yq}
  Y.~Z.~Fan, B.~Zhang and J.~Chang,
  Int.\ J.\ Mod.\ Phys.\ D {\bf 19}, 2011 (2010)
  [arXiv:1008.4646 [astro-ph.HE]].


\bibitem{Lin:2014vja}
  S.~J.~Lin, Q.~Yuan and X.~J.~Bi,
  Phys.\ Rev.\ D {\bf 91}, no. 6, 063508 (2015)
  [arXiv:1409.6248 [astro-ph.HE]].

\bibitem{Jin:2014ica}
  H.~B.~Jin, Y.~L.~Wu and Y.~F.~Zhou,
  JCAP {\bf 1509}, no. 09, 049 (2015)
  doi:10.1088/1475-7516/2015/09/049
  [arXiv:1410.0171 [hep-ph]].

\bibitem{arXiv:1407.1859}
S.~Matsumoto, S.~Mukhopadhyay and Y.~L.~S.~Tsai,
JHEP {\bf 1410}, 155 (2014)
[arXiv:1407.1859 [hep-ph]].

\bibitem{Balazs:2014jla}
  C.~Bal\'azs and T.~Li,
  Phys.\ Rev.\ D {\bf 90}, no. 5, 055026 (2014)
  [arXiv:1407.0174 [hep-ph]].

\bibitem{arXiv:1408.2223}
M.~Frank and S.~Mondal,
Phys.\ Rev.\ D {\bf 90}, no. 7, 075013 (2014)
[arXiv:1408.2223 [hep-ph]].

\bibitem{arXiv:1409.5776}
S.~Benic and B.~Radovcic,
JHEP {\bf 1501}, 143 (2015)
[arXiv:1409.5776 [hep-ph]].

\bibitem{arXiv:1501.03164}
A.~Ibarra, A.~Pierce, N.~R.~Shah and S.~Vogl,
Phys.\ Rev.\ D {\bf 91}, no. 9, 095018 (2015)
[arXiv:1501.03164 [hep-ph]].

\bibitem{arXiv:1503.01500}
M.~Garny, A.~Ibarra and S.~Vogl,
Int.\ J.\ Mod.\ Phys.\ D {\bf 24}, no. 07, 1530019 (2015)
[arXiv:1503.01500 [hep-ph]].

\bibitem{Balazs:2015boa}
  C.~Bal\'azs, T.~Li, C.~Savage and M.~White,
  Phys.\ Rev.\ D {\bf 92}, no. 12, 123520 (2015)
  doi:10.1103/PhysRevD.92.123520
  [arXiv:1505.06758 [hep-ph]].

\bibitem{arXiv:1507.02288}
A.~Butter, T.~Plehn, M.~Rauch, D.~Zerwas, S.~Henrot-Versillé and R.~Lafaye,
arXiv:1507.02288 [hep-ph].

\bibitem{Higgsportal}
For recent discussions, see J.~M.~Cline, K.~Kainulainen, P.~Scott and C.~Weniger,
  Phys.\ Rev.\ D {\bf 88}, 055025 (2013)
  [arXiv:1306.4710 [hep-ph]]
  and references therein.

\bibitem{deS.Pires:2010fu}
  C.~A.~de S.Pires, F.~S.~Queiroz and P.~S.~Rodrigues da Silva,
  Phys.\ Rev.\ D {\bf 82}, 105014 (2010)
  [arXiv:1002.4601 [hep-ph]].

\bibitem{Kumar:2013iva}
  J.~Kumar and D.~Marfatia,
  Phys.\ Rev.\ D {\bf 88}, no. 1, 014035 (2013)
  [arXiv:1305.1611 [hep-ph]].

\bibitem{Sommerfeld}
A. Sommerfeld, Annalen der Physik 403, 257 (1931).

\bibitem{NFW}
J.~F.~Navarro, C.~S.~Frenk and S.~D.~M.~White,
  Astrophys.\ J.\  {\bf 462}, 563 (1996)
  [astro-ph/9508025];
J.~F.~Navarro, C.~S.~Frenk and S.~D.~M.~White,
  Astrophys.\ J.\  {\bf 490}, 493 (1997)
  [astro-ph/9611107].

\bibitem{FR}
  A.~Alloul, N.~D.~Christensen, C.~Degrande, C.~Duhr and B.~Fuks,
  Comput.\ Phys.\ Commun.\  {\bf 185}, 2250 (2014)
  [arXiv:1310.1921 [hep-ph]].

\bibitem{MO}
  G.~Belanger, F.~Boudjema, A.~Pukhov and A.~Semenov,
  Comput.\ Phys.\ Commun.\  {\bf 185}, 960 (2014)
  [arXiv:1305.0237 [hep-ph]].

\bibitem{Moskalenko:2001ya}
  I.~V.~Moskalenko, A.~W.~Strong, J.~F.~Ormes and M.~S.~Potgieter,
  Astrophys.\ J.\  {\bf 565}, 280 (2002)
  [astro-ph/0106567].


\bibitem{Strong:2001fu}
  A.~W.~Strong and I.~V.~Moskalenko,
  Adv.\ Space Res.\  {\bf 27}, 717 (2001)
  [astro-ph/0101068].


\bibitem{Moskalenko:2002yx}
  I.~V.~Moskalenko, A.~W.~Strong, S.~G.~Mashnik and J.~F.~Ormes,
  Astrophys.\ J.\  {\bf 586}, 1050 (2003)
  [astro-ph/0210480].


\bibitem{Trotta:2010mx}
R.~Trotta, G.~Johannesson, I.~V.~Moskalenko, T.~A.~Porter, R.~R.~de Austri and A.~W.~Strong,
Astrophys.\ J.\  {\bf 729}, 106 (2011)
doi:10.1088/0004-637X/729/2/106
[arXiv:1011.0037 [astro-ph.HE]].

\bibitem{Auchettl:2011wi}
K.~Auchettl and C.~Balazs,
Astrophys.\ J.\  {\bf 749}, 184 (2012)
doi:10.1088/0004-637X/749/2/184
[arXiv:1106.4138 [astro-ph.HE]].

\bibitem{Yuan:2014pka}
  Q.~Yuan and X.~J.~Bi,
  JCAP {\bf 1503}, no. 03, 033 (2015)
  [arXiv:1408.2424 [astro-ph.HE]].


\bibitem{Giesen:2015ufa}
  G.~Giesen, M.~Boudaud, Y.~Genolini, V.~Poulin, M.~Cirelli, P.~Salati and P.~D.~Serpico,
  arXiv:1504.04276 [astro-ph.HE].


\bibitem{arXiv:1204.2795}
B.~S.~Acharya, G.~Kane and P.~Kumar,
Int.\ J.\ Mod.\ Phys.\ A {\bf 27}, 1230012 (2012)
[arXiv:1204.2795 [hep-ph]].

\bibitem{arXiv:1307.2453}
R.~Easther, R.~Galvez, O.~Ozsoy and S.~Watson,
Phys.\ Rev.\ D {\bf 89}, no. 2, 023522 (2014)
[arXiv:1307.2453 [hep-ph]].

\bibitem{arXiv:1502.05406}
G.~L.~Kane, P.~Kumar, B.~D.~Nelson and B.~Zheng,
arXiv:1502.05406 [hep-ph].

\bibitem{arXiv:1310.1915}
K.~C.~Y.~Ng, R.~Laha, S.~Campbell, S.~Horiuchi, B.~Dasgupta, K.~Murase and J.~F.~Beacom,
Phys.\ Rev.\ D {\bf 89}, no. 8, 083001 (2014)
[arXiv:1310.1915 [astro-ph.CO]].

\bibitem{arXiv:0905.2736}
Q.~Yuan, X.~J.~Bi, J.~Liu, P.~F.~Yin, J.~Zhang and S.~H.~Zhu,
JCAP {\bf 0912}, 011 (2009)
[arXiv:0905.2736 [astro-ph.HE]].

\bibitem{arXiv:0812.3202}
D.~Hooper, A.~Stebbins and K.~M.~Zurek,
Phys.\ Rev.\ D {\bf 79}, 103513 (2009)
[arXiv:0812.3202 [hep-ph]].

\bibitem{arXiv:0809.1523}
D.~R.~G.~Schleicher, S.~C.~O.~Glover, R.~Banerjee and R.~S.~Klessen,
Phys.\ Rev.\ D {\bf 79}, 023515 (2009)
[arXiv:0809.1523 [astro-ph]].

\bibitem{arXiv:0805.1244}
J.~Diemand, M.~Kuhlen, P.~Madau, M.~Zemp, B.~Moore, D.~Potter and J.~Stadel,
Nature {\bf 454}, 735 (2008)
[arXiv:0805.1244 [astro-ph]].

\bibitem{arXiv:0803.2714}
J.~I.~Read, G.~Lake, O.~Agertz and V.~P.~Debattista,
Mon.\ Not.\ Roy.\ Astron.\ Soc.\  {\bf 389}, 1041 (2008)
[arXiv:0803.2714 [astro-ph]].

\bibitem{arXiv:0902.0009}
J.~I.~Read, L.~Mayer, A.~M.~Brooks, F.~Governato and G.~Lake,
Mon.\ Not.\ Roy.\ Astron.\ Soc.\  {\bf 397}, 44 (2009)
[arXiv:0902.0009 [astro-ph.GA]].

\bibitem{arXiv:0906.5348}
C.~W.~Purcell, J.~S.~Bullock and M.~Kaplinghat,
Astrophys.\ J.\  {\bf 703}, 2275 (2009)
[arXiv:0906.5348 [astro-ph.GA]].

\bibitem{arXiv:0902.4001}
T.~Bruch, A.~H.~G.~Peter, J.~Read, L.~Baudis and G.~Lake,
Phys.\ Lett.\ B {\bf 674}, 250 (2009)
[arXiv:0902.4001 [astro-ph.HE]].

\bibitem{McCullough:2013jma}
M.~McCullough and L.~Randall,
JCAP {\bf 1310}, 058 (2013)
[arXiv:1307.4095 [hep-ph]].

\bibitem{Elor:2015bho}
G.~Elor, N.~L.~Rodd, T.~R.~Slatyer and W.~Xue,
arXiv:1511.08787 [hep-ph].
























\end{thebibliography}
\end{document}